**Unsupervised Representations Predict Popularity of Peer-Shared Artifacts in an Online**

**Learning Environment**

Renzhe Yu[1], John Scott[2], Zachary Pardos[3]

[1] *University of California, Irvine, USA*

[2] *Blackboard Inc., USA*

[3] *University of California, Berkeley, USA*

**Abstract**

        In online collaborative learning environments, students create content and construct their own knowledge through complex interactions over time. To facilitate effective social learning and inclusive participation in this context, insights are needed into the correspondence between student-contributed artifacts and their subsequent popularity among peers. In this study, we represent student artifacts by their (a) contextual action logs (b) textual content, and (c) set of instructor-specified features, and use these representations to predict artifact popularity measures. Through a mixture of predictive analysis and visual exploration, we find that the neural embedding representation, learned from contextual action logs, has the strongest predictions of popularity, ahead of instructor's knowledge, which includes academic value and creativity ratings. Because this representation can be learnt without extensive human labeling effort, it opens up possibilities for shaping more inclusive student interactions on the fly in collaboration with instructors and students alike.

        *Keywords:* student artifact, computer supported collaborative learning, representation learning, predictive modeling, popularity prediction, higher education

## Introduction

The role of social interactions has been highlighted in decades of learning theories. Recent connectivist pedagogies further promote environments where learners contribute the majority of the course material through sharing and interacting with artifacts (Anderson & Dron, 2011; Siemens, 2005). When the instructor is pedagogically de-centered from interactions, social dynamics between peers and the amount of feedback peer-created artifacts generate can significantly impact the learning experience. While online environments are well-equipped for supporting more diffuse social interactions, it poses challenges to instructors to capture fast-paced student interactions in time and take appropriate actions. Thus, it would be pedagogically meaningful to predict the popularity of student-contributed artifacts to identify the artifacts that are shaping the discourse of the class, as well as those artifacts and perspectives absent from the discourse. In this work, we investigate this issue in an online course that utilized connectivist pedagogies, allowing for artifact creation, sharing, and remixing. We represent student artifacts from three perspectives and evaluate their correspondence to the popularity of these artifacts in the class. A neural embedding model is used to synthesize the behavioral contexts of an artifact inferred from sequences of student interactions around it. A computational text-based approach is used to represent the textual content of the artifact. A set of subjective instructor ratings of academic value and creativity, plus other descriptive features, are used to form the instructor's representation of the artifact. Are instructionally valuable artifacts the popular ones? Can unsupervised approaches capture regularities of student behavior surrounding popular artifacts? Our results answer these and other questions and we suggest how this representation approach might be further developed into automated interventions to make an online class feel more inclusive to students whose artifacts might lack sufficient interaction.

## Related Work

### Peer-Shared Student Artifacts

Learning theories that emphasize the integral role of social interaction in the learning process are often traced back to Vygotsky (1978), who famously situated learning in the interaction between people mediated by tools and signs, which in turn shape how learning and world views become internalized. Later theories including *Communities of Practice* (Wenger, 1998), *Activity Theory* (Engeström, 2014) and *Connectivism* (Siemens, 2005) have further expanded the interest in the social dimensions of learning and provided an expansive theoretical foundation from which to build upon pedagogical and technological tools for nurturing social interaction for learning in digital environments.

Student artifacts are objects that students create or curate and utilize to facilitate their learning through social interactions. Discussion forum threads were the earliest and most widely seen form of student artifacts, followed by other types of social media postings (Dabbagh & Kitsantas, 2012) and more recently, with the advent of connectivist pedagogies, collaborative works created on special platforms (Håklev et al., 2017). These learner-contributed resources can not only foster student interactions but also augment instructional content, thus facilitating student discovery and engagement with materials most beneficial to their own learning and development (Rheingold et al., 2014). Along this line, the current study is contextualized in a unique course design that provides students with diverse opportunities for artifact creation, sharing, and remix within an online space.

### Predicting Artifact Popularity

As artifacts circulate across the learning community, they may be interacted by only a handful of people, or they may go viral and become visible to larger scales of audience (Shifman, 2013). This variation in what we refer to broadly as popularity can be especially pronounced in large-scale MOOCs yet present in smaller online classrooms as well. When a learner's artifacts are well-received by peers, she might experience strong feelings of social presence, which has proved to affect immediate outcomes like motivation, satisfaction and course grades and longer-term outcomes including course retention, intent to enroll and peer communication (Joksimović et al., 2015; Richardson et al., 2017). As such, when popularity varies significantly across artifacts, instructional support for students whose content fails to attract feedback from peers is crucial to sustaining healthy and inclusive online learning communities.

Toward this end, understanding what prompts users to engage with certain artifacts and being able to predict artifact popularity are of great pedagogical value. There has been much empirical work focused on predicting the popularity or attractiveness of user-generated content in broader social media research. These studies examine a wide spectrum of artifacts, including images (Gelli et al., 2015; Hu et al., 2018; Khosla et al., 2014), videos (Sharma et al., 2016), microblogs (Miao et al., 2016), among others. Features used for prediction are even more diverse, ranging from content-based encoding and user-specific attributes to sequential patterns extracted from raw log data. Most of these studies take a marketing or business perspective, focused on popularity prediction for optimizing marketing strategies and product sales.

However, less research is seen in educational contexts, where successful prediction of popularity, especially on a real-time basis, may lay the foundations of intelligent support aimed at closing popularity gaps for equity and inclusivity. Our study enriches this strand of research, and calls for more context-specific, pedagogically meaningful investigations of artifact popularity in education.

**Connectionist Representation of Learning Resources**

Digital learning environments have left the legacy of fine-grained traces of student behavior which can be utilized to open the black box of learning processes and automate pedagogically meaningful support. However, the quantity and granularity of these behavioral signals bring challenges to hand-constructed feature extraction. In this context, representation learning (i.e., learning representations from data that facilitate information extraction for prediction or other tasks (Bengio et al., 2013)) can become useful in summarizing low-level behavioral patterns. Specifically, neural networks, also advocated as a connectionist approach in cognitive sciences, have been recently introduced to featurize learning resources from their distributed interactions with learners. From content pages in college online classes (Pardos & Horodyskyj, 2019; Yu et al., 2018) and MOOCs (Pardos et al., 2017) to questions in intelligent tutoring systems (Chaplot et al., 2018; Pardos et al., 2018; Piech et al., 2015), this data-driven, unsupervised approach has been able to unveil relationships and regularities of different types of learning resources, and support recommendation and other intelligent systems through predictive analysis. Our study further examines whether connectionist representations synthesize the complicated patterns of learning resources being contributed to as well as consumed by students in collaborative learning contexts.

## Data

### SuiteC Toolkit

Our research explores data from the SuiteC education apps, a suite of online learning tools integrated with the Canvas learning management system (LMS) (Jayaprakash et al., 2017). Motivated by research on networked learning, social media and gamification, this toolkit includes three major components that encourage students to share, discuss and remix artifacts: 1) Asset Library, a repository of peer-shared artifacts with rich social networking functions; 2)

Whiteboards, a platform for real-time collaboration on remixing peer-shared artifacts; and 3) Engagement Index, gamified tool that tracks student engagement and induces social comparison.

The current study is specifically focused on the interactions that unfold around digital artifacts students share to the Asst Library, which are referred to as "assets" in SuiteC and in the remainder of this paper. Assets can be existing online or local media content like videos and photos from other web sources, or collaborative whiteboards composed by current students. When adding an asset, a student can attach supporting information including title, text description, hashtags (#hashtag) and choose a category predefined by the instructor. Like familiar social media tools such as Instagram, the Asset Library features a social media feed that orders assets by recency and allows students to view, like or add comments on peer assets (Figure 1). They can also explore the Asset Library by filtering hashtags, categories or authors. These functions combined help to shape the knowledge community by augmenting the course curriculum resources.

**Course Offering**

We investigate a fully online, for-credit course offered to residential students at a four-year public university in the United States. This 14-week course was offered in Spring 2016 and discussed topics related to literacy and learning from a global perspective. The course curriculum incorporated SuiteC-based activities including adding assets, interacting with peer assets and collaborating with peers on whiteboards. Each week a set of topical hashtags were defined, which students were required to include in their asset descriptions. This allowed others in the course to filter by hashtag to explore peer assets for a given assignment. Beyond this minimum requirement, students were free to create assets and hashtags at their will to expand their social connections. Given the overarching theme of the course, all these activities were not focused on "correct" answers. Instead, they were designed to inspire idea exchange through assets, so that students were able to learn from an interactive process similar to real literacy experiences in the digital, global world. In this case, the popularity of assets is a mirror of how (unevenly) ideas spread out in the learning community, as suggested by the literature (Rheingold et al., 2014).

With the approval from the Institutional Review Board at the university, we acquired a dataset of action logs within SuiteC, with a total of 684,095 entries. Each entry contains the metadata of a single action (aka event in the dataset) performed by a user, including event type, timestamp, associated asset/whiteboard id (if any), anonymized user id, among others. In our previous work, we leveraged all these granular records to characterize behavioral differences across students (Authors, 2019). With the focus on assets in the current study, we keep only events that are associated with student-curated assets within the course period such as creating assets, visiting peer assets and leaving comments. We further remove "ghost" assets which are associated with fewer than three events. The final dataset includes 21,402 events associated with 3,189 assets and 100 students.

## Methodology

The aim of this study is to investigate automated ways to predict asset popularity without extensive human coding. Methodologically, this involves constructing 1) automated representation of assets and 2) predictive models. In addition, we explore the characteristics of automated representations using a visualization technique, which will be described in the third subsection.

### Asset Representation

#### *Context-Based Representation: asset2vec*

In the Asset Library, a student may interact with different assets from time to time. These interactions in chronological order constitute the behavioral contexts of any asset in the sequence, which may convey properties of the asset as a word's contexts convey properties of the word in natural language. This latter scenario has been leveraged by the word2vec model to automatically compute vector representations of words (Mikolov, Chen, et al., 2013; Mikolov, Sutskever, et al., 2013). As such, we employ this model to learn vector representations for assets.

Formally, given a word sequence (e.g., sentence) in the corpus $\{w_1, w_2, \dots, w_T\}$, the word2vec model aims to represent the semantics of each word $w_t$ as a high-dimensional vector $v_{w_t}$. The rationale is that, synonyms share similar surrounding words (contexts) and their vector representations should be close in the high-dimensional space. The modeling process iteratively looks for the vector set $\{v_{w_t}\}$ that maximizes the average log probability of contextual words across all $w_t$, i.e., minimizing the following loss function:

$$J = -\frac{1}{T} \sum_{t=1}^{T} \sum_{-c \leq j \leq c, j \neq 0} \log p(w_{t+j} | w_t) \qquad (1)$$

where $c$ is the contextual window size (the number of surrounding words), and the conditional probability is computed using a softmax function over all possible words in the corpus for each given $w_t$:

$$p(w_{t+j} | w_t) = \frac{\exp\left(v_{w_{t+j}}^{c\prime} v_{w_t}\right)}{\sum_{k=1}^{W} \exp\left(v_{w_k}^{c\prime} v_{w_t}\right)} \qquad (2)$$

where $v_{w_t}$ is the vector representation of $w_t$ as mentioned before, $v_{w_{t+j}}^{c}$ is the "contextual" vector of $w_{t+j}$ when it is used in the context of a focal word, and $W$ is the corpus size.

In our analogy, each asset is treated as a word and each student's asset-related events, sorted by timestamps, constitute an asset sequence $\{A_1, A_2, \dots, A_T\}$. Once we select the vector dimension and the contextual window size, the model will compute vector representations $v_{A_t}$ for all the assets using Equations (1) and (2). The lower part of Figure 2 shows how the action (event) logs are transformed into asset sequences and how an example asset is represented as a vector from these behavioral contexts. Similar to the word2vec setting, assets sharing similar contexts in the event logs (i.e., being interacted with together with some other common assets) should be encoded to mathematically close vectors. In the remainder of this paper, we refer to this context-based representation as *asset2vec*.

### Content-Based Representation: average GloVe

In online learning communities, the content of learner artifacts has been found to play a key pedagogical role (Wise & Cui, 2018), so automated representation of assets' textual content, including titles and descriptions, may also be predictive of popularity. As most assets had short written content and their topics were not very technical, we use pre-trained word vectors from GloVe (Pennington et al., 2014) instead of training our own vectors to represent the content of assets. An extension of skip-gram (one of the model architectures of word2vec), the GloVe model also maps the semantics of each word to a vector space, but in the training process combines both prediction-based (skip-gram) and count-based (bag-of-words) approaches. Empirical work has shown that this approach performs better in extracting word meanings than previous approaches. We use word vectors that are pre-trained on Twitter texts by the GloVe research team, because the social interactions around assets are more similar in nature to Twitter communications. For each asset $A_i$ in our dataset, we concatenate its title and description and parse into a sequence of words $[w_{i1}, w_{i2}, ..., w_{iT_i}]$, where $T_i$ is the total number of words (with duplicates) in $A_i$'s title and description. The content representation of asset $A_i$ is then calculated as the average of GloVe vectors for all these words:

$$v_{A_i} = \frac{1}{T_i} \sum_{k=1}^{T_i} GloVe_{w_{ik}} \qquad (3)$$

If no word in an asset is represented in GloVe (e.g. empty text or hashtag only), we assign a zero vector. The upper part of Figure 2 illustrates this process for the example asset. Again, assets sharing semantically similar content should be encoded to close positions in the vector space. We call this approach to representing asset content *average GloVe*.

**Combined Automated Representation: ensemble**

As described above, contexts and content may both be predictive of asset popularity. So, we also construct an ensemble feature set, which concatenates *asset2vec* and *average GloVe* vectors for each asset.

**Human-Engineered Representation: instructor features**

In addition to the previous machine-learned, unsupervised representations, we also invite the course instructor to propose attributes of assets that, according to their knowledge and experience, are relevant to popularity. Some of these attributes are subjective constructs that involve manual coding, so the instructor randomly selects a subset of assets from the Asset Library, stratified by assigned weekly hashtags and number of views, and codes this sample. Putting all the instructor-identified attributes together, we construct the *instructor features* listed below. This feature set can provide a baseline to compare with the foregoing automated features.

- *Academic Value*: How effectively the asset demonstrates a student's understanding of a particular course concept or reading, including establishing a clear relationship between the concept and what is being represented in the asset (coded on a 5-point Likert scale)*Creativity*: A social measure of the affective impact of the asset, either aesthetically in terms of its design or in the content used to demonstrate or elaborate on a course concept (coded on a 5-point Likert scale)

- *Day from assignment*: Time difference between the day of focal asset being added and the day of its associated hashtag being assigned

- *Title length*: Word count of the asset title

- *Description length*: Word count of the asset description

- *Type*: Asset type, based on its format and/or its function in the course, including simple asset (single artifact created and shared by a student), collaborative whiteboard (multimedia artifact containing multiple assets created by two or more peers synchronously or asynchronously), solo whiteboard (multimedia artifact containing multiple assets created by one student) and curated asset (single artifact searched and shared by one student from the web)

In Figure 2, the middle part presents the numerical values of these features for the example asset.

**Prediction Models**

One major theme of this study is exploring what asset features encode signals of popularity. In our course, popularity is defined as the number of different students who interacted with an asset throughout the course. Because of our small sample size, we employ two relatively simple model frameworks to predict popularity.

*Regression Model*

Because popularity is a count variable, we assume that it follows a Poisson distribution and employ a Poisson regression model:

$$\log(y) = wx \tag{4}$$

where $w$ is the weight vector, $x$ is the feature set (asset representation). The model is implemented as a simple "neural network" without hidden layer in the Keras package (Chollet, 2015). An exponential activation function is imposed before the final output. Poisson loss is used in the training process:

$$J(w) = \frac{1}{N} \sum_{i=1}^{N} (\hat{y_i} - y_i \cdot \log(\hat{y_i})) \tag{5}$$

where $\hat{y}$ denotes the predicted popularity.

### Neural Network Model

Slightly more complicated is a simple feed-forward neural network with one hidden layer. The input layer receives different asset representations, which are projected to the hidden layer. After a hyperbolic tangent transformation, the hidden outputs are used to calculate the single output neuron before it is activated through an exponential function to produce the final prediction of popularity. The same Poisson loss function (Equation 5) is used in the training process. Figure 3 illustrates the entire model architecture, which is also implemented in Keras.

### Visualization of Vector Representations

Context-based and content-based asset representations are both high-dimensional and therefore challenging to interpret in their raw formats. To visualize how these assets are distributed in the vector space, we use Barnes-Hut t-SNE algorithm to reduce the dimensionality to two (Maaten, 2014). Essentially, this algorithm is a non-linear manifold projection which retains the relative positions of high-dimensional vectors in lower-dimensional space. These projected asset vectors are depicted on a 2-dimensional plane by an interactive d3 scatter plot tool developed by the last author's research lab (blinded). This tool also supports coloring data points based on their attributes, so that users can get a better view of the relationship between vector representations and outcomes of interest, or popularity in our case.

<div align="center">

**Results**

</div>

## Asset Representations

To compare different representations of assets, we confine our analysis to the subset of assets that are both kept in our final dataset of event sequences and coded by the instructor. This selection results in a total of 277 assets. Note that, however, *asset2vec* and *average GloVe* representations are trained from our final dataset of event sequences which includes 3,189 assets.

Informed by prior research in similar contexts (Pardos & Horodyskyj, 2019), we choose an array of candidate hyperparameter values to train *asset2vec* and *average GloVe* representations. After iterating through these values and testing our prediction models, we decide on a vector size of 50 for both models and a window size of 3 for *asset2vec*, which in general lead to the highest prediction performance. For the neural network model, a hidden layer of 8 units is employed.

For *instructor features*, Table 1 reports their descriptive statistics. The upper panel shows that the manually coded variables (academic value and creativity) both have balanced, symmetric distributions. Without a second human rater, this validates to some extent the instructor's coding. Students on average begin working on asset sharing five days after the assignment date and the median is six days. Given the seven-day (weekly) cycle of asset assignments, this suggests a habit of slight procrastination among many students. Most student-shared assets are brief in their title and description, while there are a small proportion of lengthier ones. The lower panel suggests that most assets are simple assets from external sources or finished collaborative whiteboards, which are the two major sources of assets by design.

Furthermore, the first five rows and columns of Figure 4 show the rank correlations (Spearman's ρ) between these features (except for asset type). The most notable relationship lies between academic value and creativity, with a coefficient of nearly 0.7. Day count from assignment shows consistent negative correlations with all other features. In other words, procrastination is associated with more hasty, low-quality work. On the other hand, lengthy assets receive higher scores on academic value and creativity.

**Predicting Popularity**

Figure 6 shows the popularity distribution of these 277 assets in our analysis. The shape resembles that of a Poisson distribution, which validates our choice of Poisson-based predictive models. The last row and the last column of Figure 4 along with Figure 5 shed light on the relationship between popularity and *instructor features*, which is moderate at best ($\rho = -0.317, p < .001$ between popularity and day from assignment). Note that neither academic value nor creativity is strongly correlated with popularity.

In our formal prediction task, *instructor features* are standardized to Z scores, and the machine learnt asset vectors of *asset2vec* and *average GloVe* are all normalized to unit length (before the two being concatenated to form *ensemble* representations). Five-fold cross-validation is performed and, within each fold, 20% of the training data are used as the validation set in the training process to avoid overfitting. Root mean square error (RMSE) is calculated to evaluate the results.

Figure 7 plots the performance of these different models. In the baseline scenario, the mean popularity of training data serves as the predicted value for every test sample in each fold. Overall, the neural network framework does not lead to better prediction performance than Poisson regression, so the following discussion is focused on results from the latter. Consistent with the correlational analysis, *instructor features* predict popularity only moderately better than the baseline ($|\Delta_{RMSE}| = 0.277, t = 3.352, p = .001$), suggesting limited human knowledge about nuances of student interactions contributing to popularity. Among the data-driven representations, *asset2vec* outperforms the *instructor features* to a greater extent ($|\Delta_{RMSE}| = 0.740, t = 5.786, p < .001$), while *average GloVe* seems unable to encode latent features of popularity. In the ensemble scenario, these two representations do not add up to better performance; instead, *asset2vec* is held back by *average GloVe*.

In order to delve deeper into the error distribution across individual assets, we estimate for each asset representation a kernel density function of absolute prediction errors from Poisson regression. Figure 8 shows that across all these representations, the errors have long-tailed distributions, where most predicted values are not distant from the actual popularity measure. *Asset2vec* has the highest peak to the left end, indicating a high proportion of accurate predictions. Meanwhile, *average GloVe* has the flattest error distribution of all five curves. This figure reaffirms the conclusions from Figure 7. Although both data-driven representations involve low-level featurization of assets, only contextual signals are indicative of popularity. Regularities of content, which are often good predictors of social media's popularity (Gelli et al., 2015; Khosla et al., 2014), fail to do well in our context. This leads us to examine the representations through exploratory visual analysis in the following subsection.

**Visual Analysis**

Following the methodology described above, Figures 8 and 9 present the visualizations of *asset2vec* and *average GloVe* representations of all the 3,189 assets in the final dataset. Each point represents an asset in the final event sequences and their locations on these 2-D planes reflect their original positionality in the high dimensional space. To examine how each of the representations encodes signals of popularity, we color the points by logarithm of popularity measures.

Figure 9 presents obvious structures in terms of both the shape and coloring. The least popular assets are sparser and present the upper left quadrant. As the popularity increases, the assets become more compact and the most popular assets are clustered together in the bottom right quadrant. Because *asset2vec* representations extract contexts of assets being accessed by students, this 2D reduction indicates that popularity is a top factor that can be differentiated by behavioral contexts. In Figure 10, by contrast, there is an absence of observable regularities. Assets exhibit unusual positionality, with several spherical clusters of different sizes and scattered points all over the space. When we scrutinize the original content, we find that assets in the spheres only contain functional words (e.g., "Week 1") and/or hashtags (not valid words). This complex nature of assets adds to the difficulty in content-based representation and might lead to overfitting in prediction models. However, even if we look solely at scattered points which make up a more natural shape, assets are not clearly grouping based on popularity. These irregular patterns might be attributed to the overall shortness of asset descriptions as seen in Table 1. Both visuals resonate with the prediction results, illustrating how differently the two data-driven representations capture signals of popularity.

**Discussions and Conclusion**

This study explores various features and conditions of student artifacts that may predict their relative popularity in the context of an online course. A main objective here is to investigate novel data-driven, unsupervised features that may potentially facilitate in-situ pedagogical interventions. Taking advantage of the enhanced features for artifact creation and sharing enabled in the SuiteC toolkit, we construct three different representations of assets (artifacts within SuiteC) utilizing various sources of information from low-level student behavior to the instructor's professional knowledge, and compare how well they capture signals of popularity. We find that *asset2vec* representation, which encodes behavioral contexts of assets being interacted with by students, is the most predictive of popularity among the three, even more than the instructor's expert knowledge. By contrast, *average GloVe* which represents the textual content of assets, hardly captures more signals of popularity than a naïve baseline. From the visualization of these representations, we further validate that assets with similar behavioral contexts are projected to the same region in the vector space of *asset2vec*, and that the positionality of these regions shows clear relationship with asset popularity.

The strong predictive power of *asset2vec* leads us to envision the potential pedagogical utility of such vector embedding techniques in SuiteC and other similar learning communities. While certain instructional contexts may focus on surfacing "best" responses from students and promoting the visibility and hence popularity of those artifacts, oftentimes the instruction needs to prioritize the balance in popularity such that all student can have their voices heard and feel a sense of inclusion and belonging, especially when the learner population is heterogeneous. As such, the instructors or facilitators of these learning environments may benefit from receiving real-time forecast of the popularity of certain content or authors in order to foster a more inclusive environment, or a network where connections do not cluster around "super nodes." One specific possibility is vector-based pairing. If the semantics of the learned vector space have the same mathematical properties as other word2vec style embeddings (Mikolov, Sutskever, et al., 2013), then an artifact predicted to be less popular say, $A_1$) could find its "partner" ($A_2$) such that $v_{A_1} + v_{A_2} = v_B$, where $B$ is some "beacon" of popular artifact. This may allow for real-time pairing or grouping of such artifacts or their authors simply based on students' action sequences.

For students, such mechanics can be further weaved into intelligent prompts for self-reflection and social comparison regarding their own artifact content, or recommendations of collaborators. For instructors, this architecture can help them make decisions in a timely fashion to share complementary artifacts to students who runs the risk of being marginalized in the community. Importantly, given the discrepancies we observe between what the instructors sees as pedagogically valuable and creative and what garners much attention from the class, instructors may evaluate the tradeoff between the two, using their expert knowledge in tandem with the automated paring and grouping, to make more responsible recommendations. In short, given the velocity of artifact sharing and the complexity of social behavioral structures, our unsupervised representation becomes promising in contributing to more inclusive learning communities through a human-algorithm collaboration paradigm. Recent research has touched upon this inspiration in hopes of communicating misconception information in intelligent tutoring systems (Pardos et al., 2018). Due to logistic constraints of the current research collaboration, we are not able to delve deeper into this idea, but it remains an important line of future work. One limitation of this study is the small sample size compared to typical scenarios where connectionist representations are investigated to encode behavioral contexts. While the *asset2vec* representation succeeds in capturing more signals of popularity than human-engineered features do in our case, the utility of this approach needs further validation in social learning scenarios of larger scale. In terms of methodology, our current content-based representation only leverages the plain texts of assets, but these socially contributed artifacts contain a mixture of natural language, non-word hashtags, and most importantly, visual content components. Prior work has investigated representation of social media items via a mixture of low-level computer vision features and textual information (Chaplot et al., 2018; Khosla et al., 2014; Sharma et al., 2016).

As such, a future direction would be to further explore "multimodal" representations of assets that combine more perspectives on content in order to more comprehensively represent their properties and pedagogically utility. Towards the goal of building in-situ support systems, additional future work may also incorporate student-centered, qualitative approaches to gain deeper insights into how students create and engage with assets and what recommendation strategies would be more useful for them.

**Table 1**

*Descriptive Statistics of instructor features (N= 277)*

*Continuous Variables*

| Variable | Mean | SD | Min | Median | Max |
|----------|------|------|------|--------|------|
| *Acad* | 3.159 | 1.19 | 1 | 3 | 5 |
| *Creativity* | 3.134 | 1.21 | 1 | 3 | 5 |
| *DayAsgmt* | 4.953 | 3.758 | -5 | 6 | 43 |
| *TitleLen* | 2.747 | 1.718 | 1 | 2 | 15 |
| *DescLen* | 44.292 | 62.253 | 0 | 1 | 344 |

*Categorical Variables*

| Variable | Count | Percentage |
|----------|-------|------------|
| *Type* | | |
| CollabWB | 105 | 37.91 |
| Asset | 98 | 35.38 |
| SoloWB | 57 | 20.58 |
| Curated | 17 | 6.14 |

**Figure 1**

*Gateway Page of the Asset Library Displaying the Asset Feed*

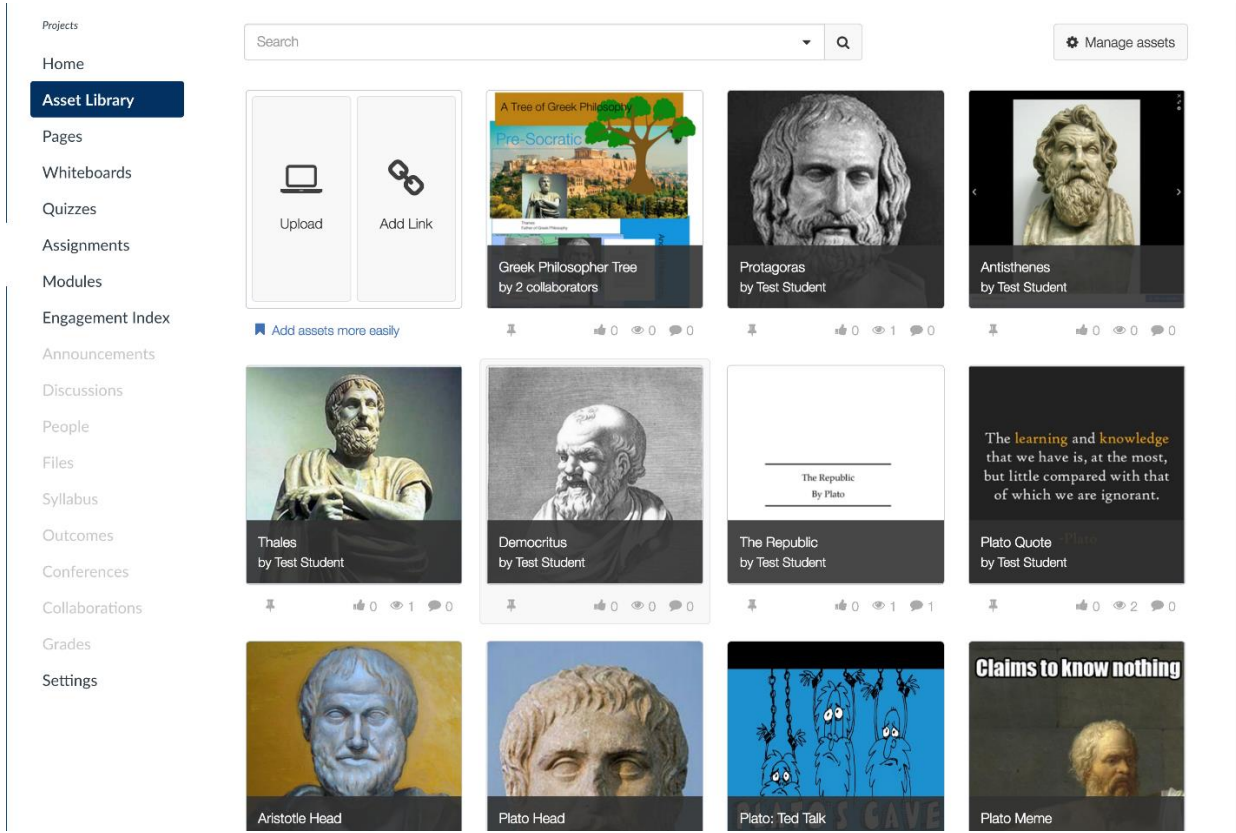

**Figure 2**

*Illustration of Asset Representation Approaches*

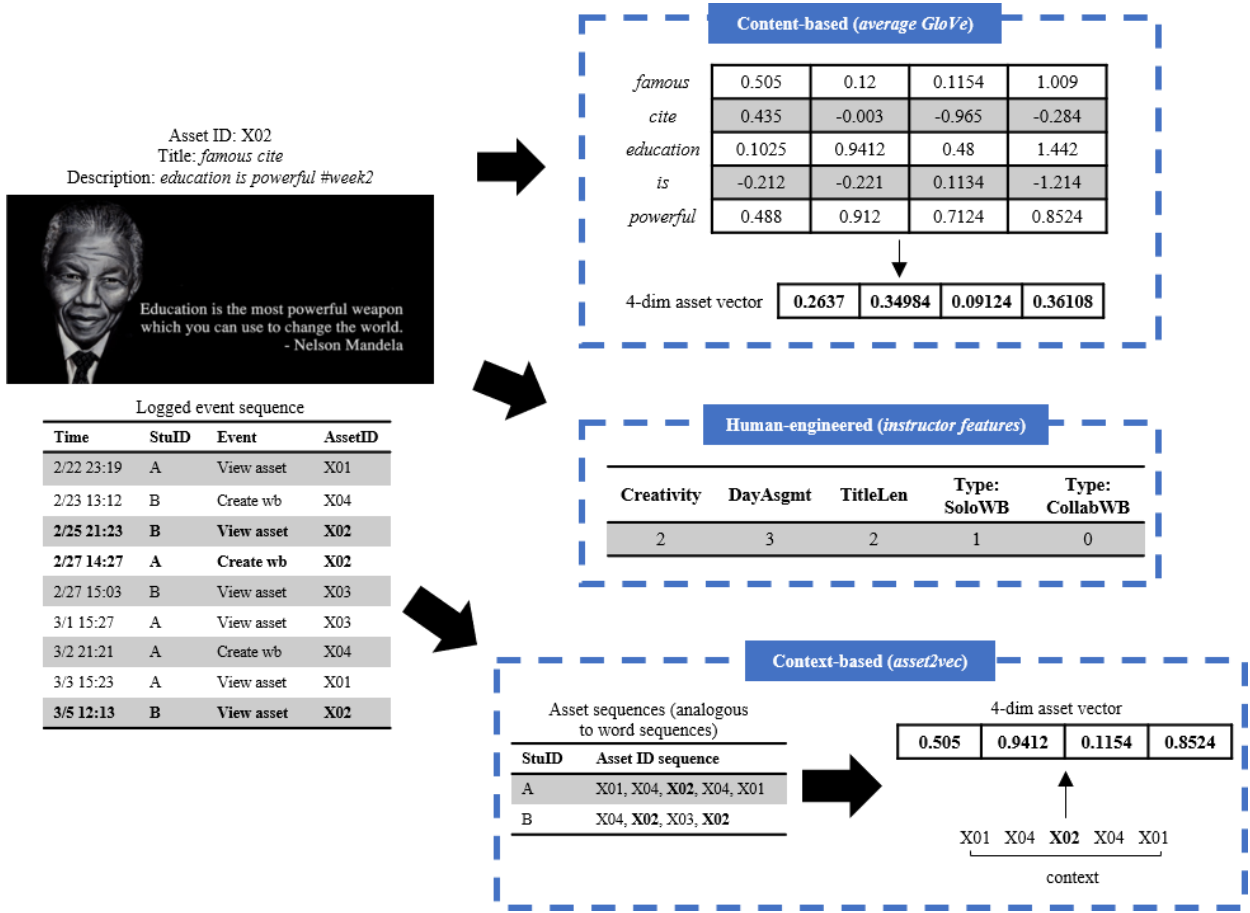

**Figure 3**

*Neural Network Architecture for Predicting Popularity*

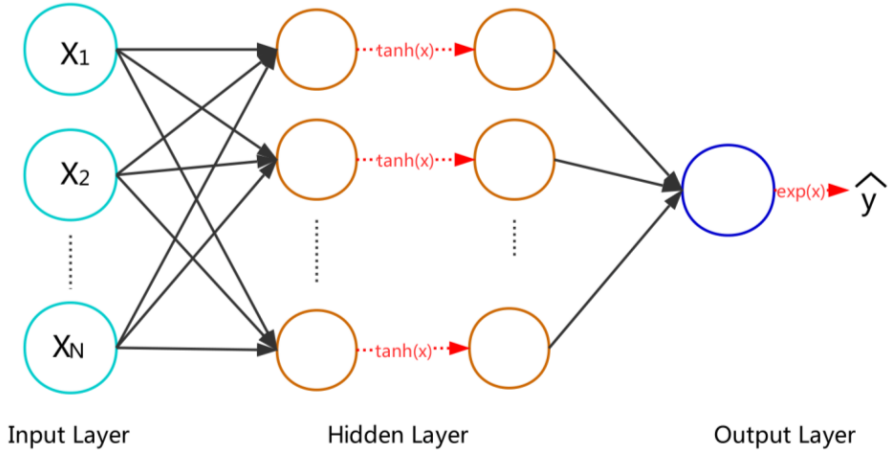

**Figure 4**

*Rank Correlation Between Asset Popularity and instructor features*

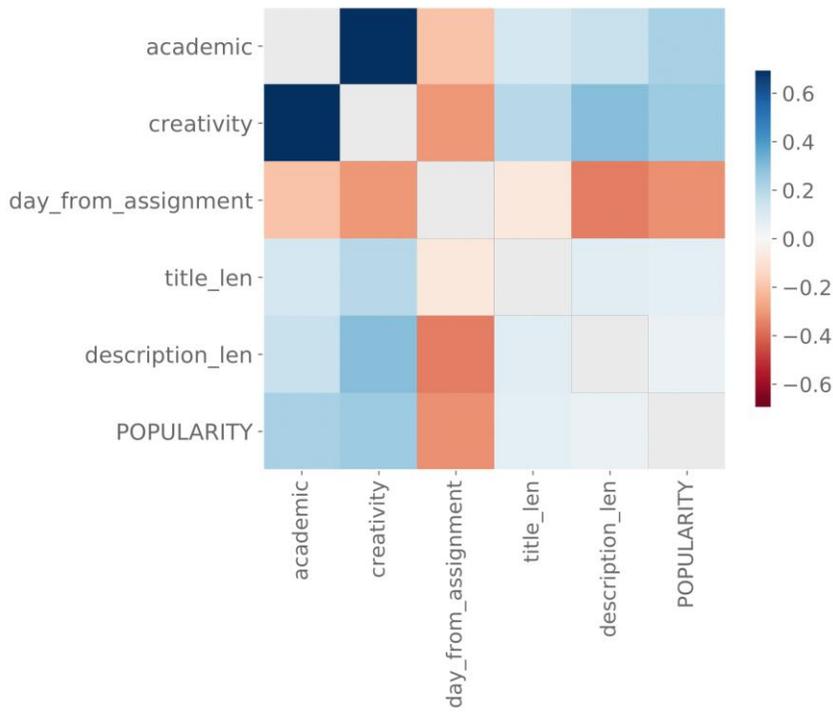

**Figure 5**

*Popularity Distribution by Asset Type*

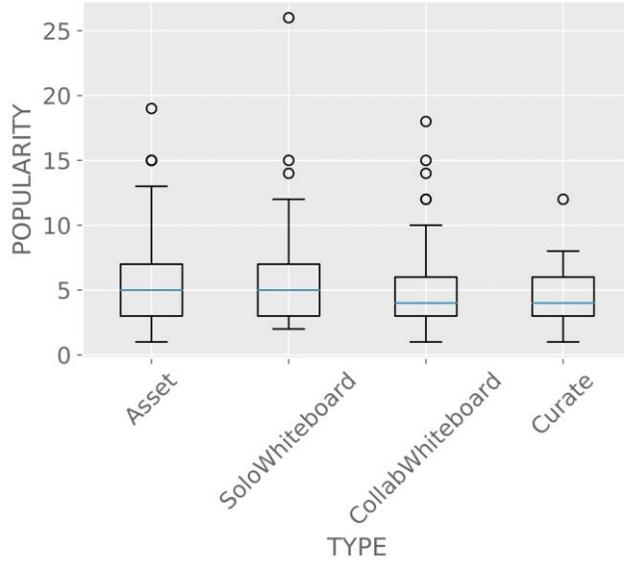

**Figure 6**

*Distribution of Asset Popularity (N= 277)*

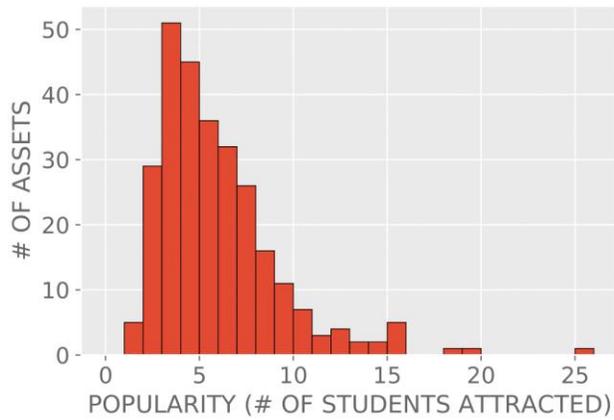

**Figure 7**

*Performance of Popularity Predictions Across Asset Representations and Model Frameworks*

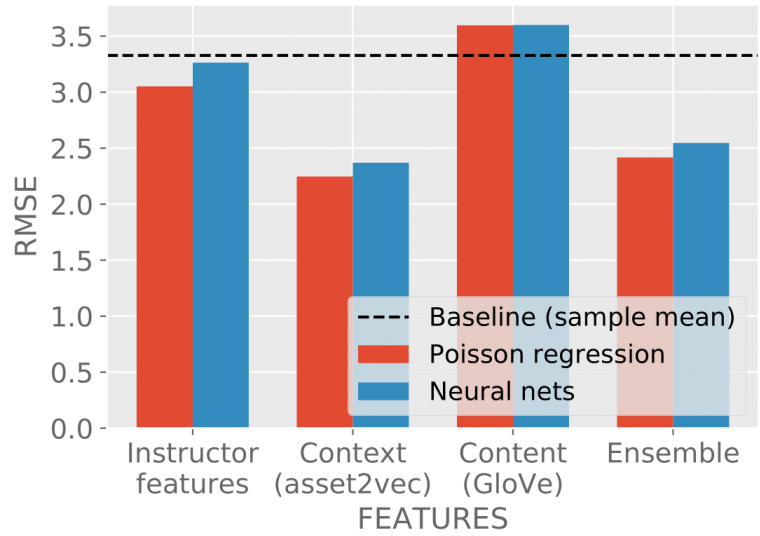

**Figure 8**

*Estimated Density Function of Absolute Prediction Error from Poisson Regression Models*

*Using Different Asset Representations*

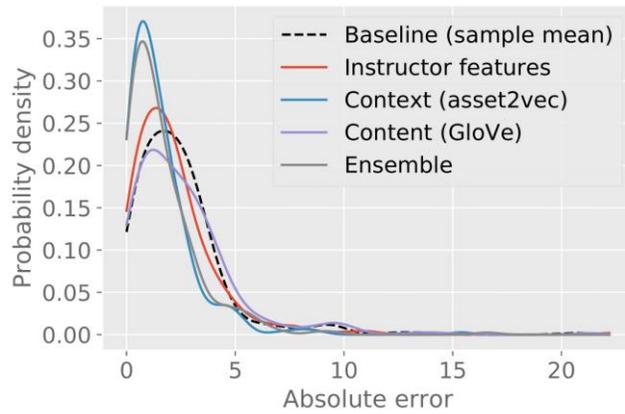

**Figure 9**

*Visualization of asset2vec*

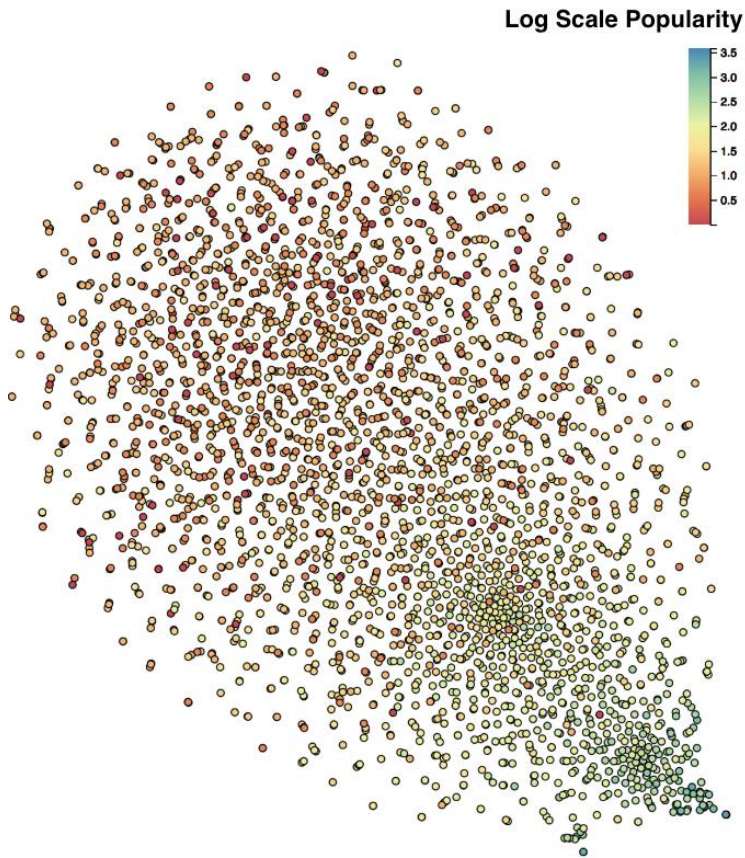

**Figure 10**

*Visualization of average GloVe*

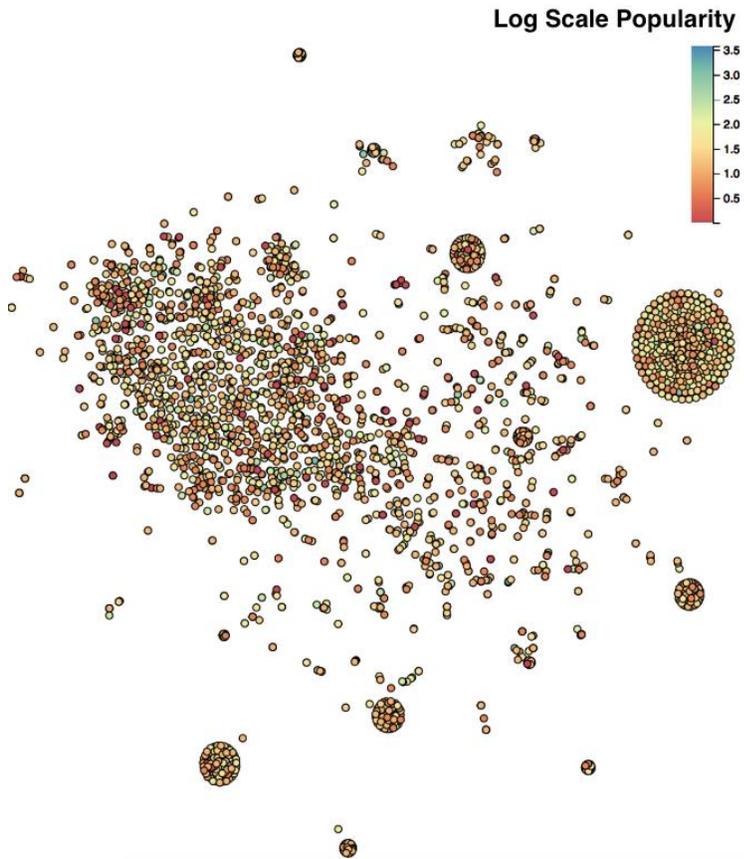